# The Conflict between Economic Development and Planetary Ecosystem in the Context of Sustainable Development

CORINA-MARIA ENE, ANDA GHEORGHIU,
CRISTINA BURGHELEA, ANCA GHEORGHIU
Hyperion University, ROMANIA
corina.maria.ene@gmail.com, andagheorghiu@yahoo.com, crystachy@yahoo.com, anca.gheorghiu@gmail.com,
www.hyperionline.ro

*Abstract:* *The green area of economy is the key of healthy living. It is necessary to convene economic and ecologic framework to establish a market attentive to drastic reduction of emissions damaging our climate and landscapes in rural areas, to the protection of biological diversity of the planet, to stop producing nuclear waste, etc.*
   *This paper tries to demonstrate human concern for a waste recycling economy that will provide new jobs, will create economic and social stability and will ensure a healthier and cleaner environment.*
   *Green Economy and its support system (planetary ecosystem) won't be in conflict anymore. Green Economy will be able to support economic progress for future.*

## 1. Introduction

In the early '70s the world was shocked by a series of forecasts anticipating that the future will be of abundance, but it will be suffocating wealth coexisting with unmet needs and pockets of poverty.

In 2008, the year when the current global crisis reached its climax, researcher Peter A. Victor wrote, that the *Limits to Growth* has had a huge impact on how we still think of environmental issues. The models in the book were meant to be taken as forecasts "only in the most limited sense of the word" as they wrote. [1]

Notice that economic growth would be the best way of human growth that proved to be an illusion as long as no consideration was given to unforeseen consequences. The myth of economic growth has been shaken by deep crises: ecological, social, economic and human.

## 2. Orientation towards green economy era

The ecological crisis occurs because of pollution, it is therefore an uncontrolled growth resulting from excessive growth of industrialization. Important question is: what should we do? Let economic growth slow to avoid damaging ecological system or do we need to find effective remedies to avoid such consequences?

The solution is green economy. But what is green economy? Green Economy is the modern world economy, labour markets, human needs, raw materials and how they all intertwine harmoniously so as to create symmetry between human needs and environmental needs. Green Economy emphasizes the use and reuse of substances/objects. Not on product quality only but less on quantitative side: recovery, recycling, reuse and less on the accumulation of wealth or material goods[2]. Green is our future: a healthy economy that we should find ourselves at a certain time (See Figure 1).

Green economy requires rethinking about current economic activity, taking into consideration of other economic processes than only those used so far, resulting an economic system harmonizing with natural systems that operates on principles of sailing boats, driven by

---

ecological processes, not on non-renewable natural resources or depletion.

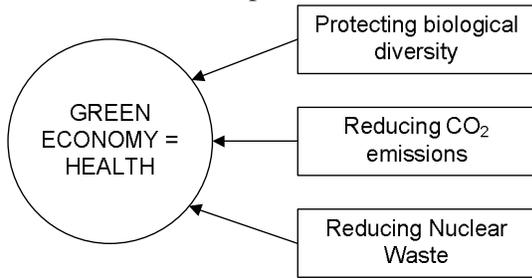

**Figure 1** - Green economy health

Keywords covering the size of a green economy are: the use of intrinsic values in nature, quality, and existence of natural flows in the workings of the economy. It must be based on natural processes, such as renewable natural water cycles, local materials, local supply networks etc., waste/food waste is to be recycled more so if such can be made into new products. Diversity can be applied at all levels. It involves creativity and human development. It requires the reconstruction of production units on ecological bases (See Figure 2).

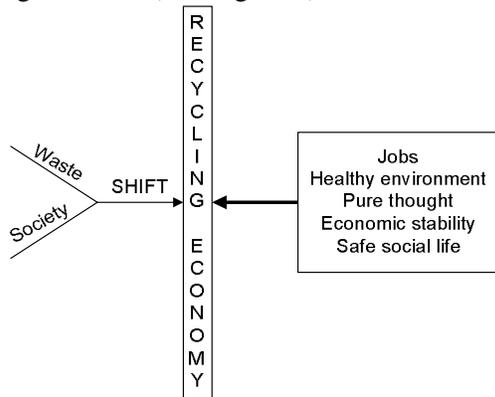

**Figure 2** - Safety shift towards a recycling economy

Green economy needs "pioneer companies" that can change the hostile economic landscape of the present, prepare to attract other companies to produce green products. Only the practice of creating a home for programming at the field for ecological alternatives can produce positive results.

It is necessary to learn how to adapt to new conditions? In this regard, education plays a key role in shaping the culture for quality of life in accordance with environmental conditions.

## 3. Viability of an economic sustainable society

Life of our planet is threatened by various phenomena specific to modern civilization; sometimes, scientific and technological progress is harmful or even fatal to our planet. The problem of pollution is extremely dangerous. It fails to distinguish apparent: air, water, and soil pollution that threatens the future of the planet, affecting all nations. Glass, paper and plastics have to be recycled. There are washing machines that work without detergents. Riding a bicycle is healthier than going by car. The use of transport, especially electric ones, means reducing pollution in large cities. Deforestation and watershed degradation lead to desertification (See Figure 3).

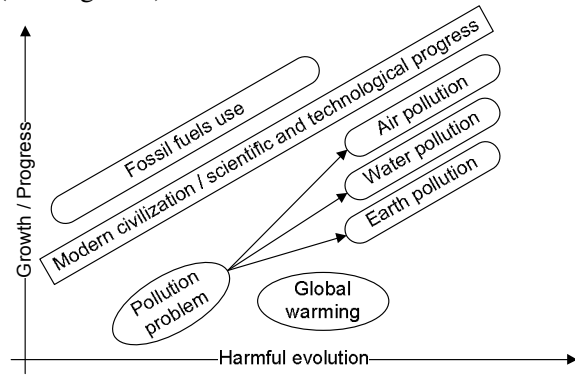

**Figure 3** - Harmful evolution of modern civilization

A sustainable society is one that is able to satisfy its needs without jeopardizing the future of forthcoming generations. It depends on the responsibility of each generation to ensure the next generation can enjoy the same natural and economic wealth. In recent decades, the economies of most developed countries have had a primary resource: the fossil fuel.

The problem of air pollution, melting glaciers and irreversible threat of global warming shows that the development of society is far from being viable. It is moving quickly towards its extinction. Clearly, an economy that depends on using polluting substance to produce food is not a viable economy. It leads quickly to climate change, since it is based on the support of massive deforestation to provide fuel for the process. We must provide a development environment for the benefit of future generations (See Figure 4).

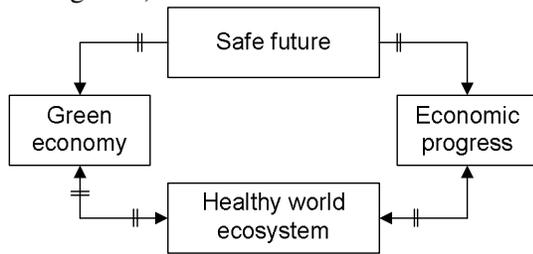

**Figure 4 -** Safe economic progress

Rapid increase in population pressures especially the third world countries adversely. It is a major threat to global water system. Future population predictions vary between 7 and 16 billion (see Figure 5). It should be reasonably coordinated and people should understand the connection between the average number of persons in a family and their life quality.

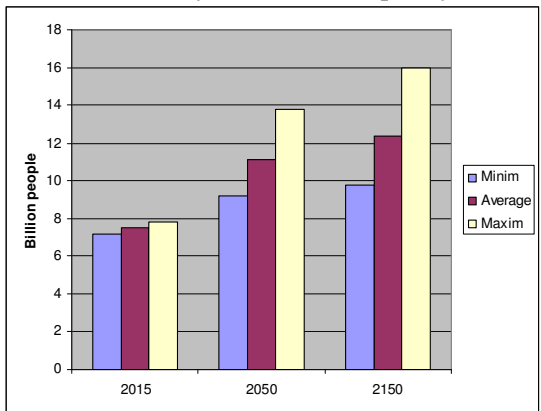

**Figure 5 -** Forecast of future population
Source: author's proposal

Population uses more resources than it needs; much of the resources are taken from the environment and converted to waste. The quantity of waste has outgrown the capacity of recycling, forming an unknown phenomenon until the coming out of man - the polluter. The way they used energy resources, natural and anthropogenic processes has accelerated.

The future of our planet now seems irrevocably urban, so we need to ensure that city life is healthy, fair and fits the concept of sustainable development. First inhabitants of towns had problems because of new urban housing. Congestion caused Tuberculosis, typhus, bubonic plague and smallpox appeared in many of the first cities. Often, people lived shorter lives, on average, than their ancestors, living in the countryside. In poor urban areas of countries, people bear a "double burden" of disease and poverty, which is an even more daunting challenge to human health on an urbanized planet.[3]

Not long ago, fossil fuels were cheap and abundantly available and therefore, they have been used very inefficiently. The rapid increase in oil prices, natural gas and coal are a signal that the time when fossil fuels were cheap is over. Competition for fossil fuels, tense international relations and such a trend is likely to intensify in the future.

International Energy Agency (IEA) has forecast that renewable energy will account for 13% of global energy in the range from 2005 to 2030. If implemented policies which are now at the stage of analysis, this percentage could reach 17%, and renewable could provide 29% of gross energy in this interval.[4]

At present, hydroelectric energy is the world's largest source of renewable energy (over 80 % of the world's renewable energy is hydroelectric). Unfortunately, according to the International Energy Agency (IEA), fossil fuels currently provide 81 % of the world's primary energy supplies. Total world electricity still depends primarily on fossil fuels but to a somewhat lesser extent. Hydro and other renewable energy sources account for 18.2 % of the world's total electricity needs. Fossil fuels still account for well over half of the world's electricity supplies – 66.6 %, while nuclear energy supplies 15.2 % of the world's electricity.

Substitution of fossil fuels involves a dual strategy: to reduce energy consumption through energy efficiency and meeting the needs uncovered a large part of renewable sources. IEA estimated that you will need to invest $45 trillion or an average, 1% of the current global economic output by 2050 for the world to

---

dispense oil and reduce $CO_2$ emissions by half[5] (See Figure 6).

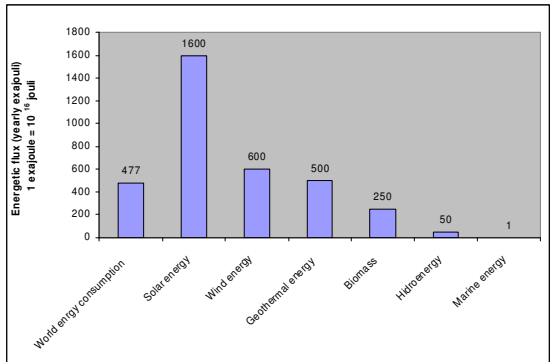

**Figure 6 -** World energy consumption in 2009 and annual production potential of renewable energy with current technologies
Source: The World-watch Institute „World state – About global warming 2009", p.182

Efficient energy services using natural energy flows will protect the global climate, will enhance the economy, will create millions of new jobs, and will help developing countries to reduce poverty. Hence, the personal and social security will increase in all countries; improve both human health and the ecosystem.

Cheap and readily available energy period favoured the construction of modern buildings rather in opposition to nature than in harmony with it.

Recent researches reveal a cruel truth: the average annual increase of $CO_2$ concentration is now 2.1 ppm/annum, with 40% more than ten years ago. Decreasing the level of $CO_2$ at an acceptable level of 350 ppm is the crusade essential for sustainable development, in an era of irrational climate and economic crisis. The most dramatic annual carbon emission in the past decades happened in America.

A zero-energy waste buildings and zero-carbon is a building that produces its entire local energy needs from renewable sources, and does not emit $CO_2$ at all. Passive solar heating and heat storage in buildings can significantly reduce the need for additional heating. In many locations, thermo solar panels can heat hot water and good profitability. Photovoltaic cells (PV) can be integrated into the roof and even the facades of buildings, typically cheaper than weather protection materials.

# 4. Sustainable development - Romanian case study

In order to improve the management of natural resources and to prevent their over-exploitation, and recognition of the value of ecosystem services, Romania must recognize that ecosystems provide a wide variety of services. Determining clear arguments justifying the financial and human efforts for the conservation of natural ecosystems and maintaining biodiversity is very important. One way to identify such arguments is to investigate what it depends on natural ecosystems, biodiversity and what we lose as a consequence of ecosystems and species extinction.

In Romania, wind aggregates can be installed in areas where the average wind speed is at least equal to 4m/s, the standard being that of 10 meters above the ground. Black Sea shore has a high energy potential because in this area, the annual average of wind speed exceeds the threshold. In Romania, one can install wind turbines with a capacity of up to 14,000 MW, which means an intake of about 23,000 GWh/year of electricity per year.

Solar energy is used mainly for the production of hot water in individual homes. There could be replace about 50% of needful hot water or 15% of the energy used for the heating current if it were fully exploit the full potential of solar power in Romania. In the solar weather conditions in Romania, a solar heat collector working under normal conditions of safety for a period from March to October, has a yield ranging between 40% and 90%.

Another source of renewable energy in Romania that has a great potential is water, which is used in a ratio of 48%. Biomass fuel is mainly used in the countryside for heating homes and water. Exploiting the potential of biomass requires the use of logging residues, sawdust and other wood waste, agricultural waste from grain or corn stalks, vine plant debris and urban waste residues.

The regional distribution of renewable energy potential:

• Danube Delta - solar energy
• Dobrogea - solar and wind
• Moldova - micro-hydro, wind, biomass
• Carpathian - micro-hydro and biomass
• Transylvania - potential for micro-hydro
• Western Plain - geothermal energy
• Sub-Carpathian - biomass and micro-hydro
• Southern Plain - biomass, geothermal, solar

Now, Romania produces about 18% of its energy from renewable sources. The European Commission has set the 2020 target to 24%, with intermediate targets: in 2012 to 19.04%, in 2014 to19.66%, in 2016 to 20.59% and in 2018 to 21.83%. At the EU level only 8.5% of energy derives now from renewable sources, while by 2020 the aggregate target is 20%[6] (See Figure 7).

Between 2007 and 2013, Romania must reduce the existing gap with other EU Member States on environmental infrastructure, both quantitatively and qualitatively, through the development of efficient public services in the field, through implementing the concept of sustainable development and the principle of "polluter pays".

The strategies and national needs meet the EU Environmental Guidelines for Sustainable Development Strategy (renewed) and aimed at attaining specific objectives.

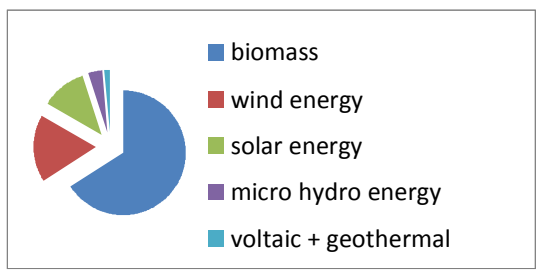

**Figure 7 -** Romanian capacity in green energy field
**Source:** Press communicate from Ministry of Environment and Forests, accessed from www.mmediu.ro at February 18, 2010

To achieve these objectives, Romania obtained transition periods for compliance with the European Acquis for the collection of municipal wastewater discharge by 2015 for 263 towns of more than 10,000 inhabitants; for 2346 towns between 2000 and 10000 inhabitants by 2018; and by 2015 for improving the quality of drinking water.

By 2013 the Program (approved in 2005) to eliminate discharges, emissions and losses of hazardous substances into the aquatic environment to prevent pollution of inland surface water resources, coastal marine ecosystems and groundwater (Directive 2006/11/EC) will be applied.

Romania has been granted transition periods to comply with EU directives:

- municipal landfill until– 2017;
- temporary storage of hazardous waste by 2009;
- storage of hazardous industrial waste by 2013;
- Compliance with emission limits (sulphur dioxide, nitrogen oxides and dust) to comply with EU directives on reducing emissions from large combustion plants.

An important objective is the preservation of natural heritage and biodiversity in protected areas management, including the implementation of Natura 2000.

Natura 2000 sites represent 17.84% of the country's surface, including 273 sites of Community importance (13.21% of the surface). The National Agency for Protected Areas and Biodiversity Conservation, which became operational in 2008, provides overall coordination of development and implementation of management plans for each of the sites designated for protection.

Attaining the average key parameters of responsible management of natural resources can be achieved by 2020, to the extent of the funds necessary to cover the financing of water management and water treatment, according to the pledges made by the Treaty of Accession to the European Union.

Regarding the issue of improving air quality –it will continue the rehabilitation of central heating systems, leading to respecting the limits of $SO_2$ emissions and contaminants prescribed by EU Directives.

In regard to integrated waste management, it will move gradually from the selective collection of waste storage to recovery in a greater proportion of recyclable waste, including by composting the organic waste, and the use of organic warehouses only for urban areas. Such conditions are expected to be reached in 2030 by approaching significantly by the environmental performances of other EU member states. Romania is in line with EU requirements and standards for water and wastewater management, according to the Preliminary Projections of River Basin Management Plan.

---

[6] Thies Frauke, Greenpeace expert for UE, www.wall-street.ro, accessed at 19.02.2010.

The EU Common Agricultural Policy will also play a significant role in Romania's sustainable development. After 2013, EU policies in this sector will be focused on three main objectives:
- To ensure food supply to guarantee sufficient amounts of food markets, food quality and food diversity.
- To ensure a proper and sustainable management of natural resources.
- To provide mutual support throughout the EU states.

Increasing energy costs and the need to reduce emissions of greenhouse gases represent major constraints for increased production. The impact of climate change will eliminate the option of holding large areas of additional land. Climate changes will also cause, water shortages and droughts, which in turn will act as a deterrent against increasing production.

European agriculture and from elsewhere will have to produce more food using less land, water and energy. European farmers and the CAP must demonstrate that they have answers to the challenges of the 21st century and that they themselves are part of the solution, not a cause.

## 5. Conclusions

Green Economy is the solution for a healthy, prosperous, clean life. Economic policies should establish an environmental framework for the market clearly aimed at: reducing our emissions that are drastically deteriorating climate, natural landscapes in rural areas, protecting diversity of biological world, to stop producing nuclear waste.

The solutions offered by green economics are not yet implemented because people still hold onto selfish mentalities, without realising that it is in their own interest to respect ecological rules.

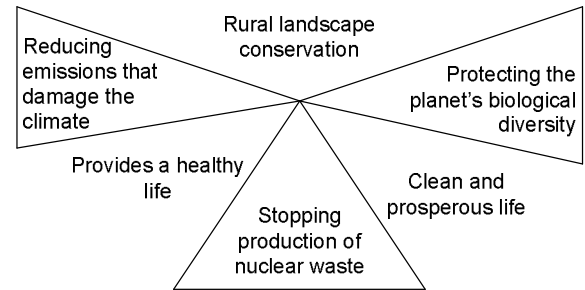

**Figure 8 -** Green economy solutions

Humans still want to indulge in immediate success and do not think of a successful future that can actually offer a chance for better life (See figure 8).